\documentclass{skaox2006}
\usepackage{graphicx}
\begin{document}
   \title{Limits on CMBP $B$--Mode Measurements by  Galactic Synchrotron Observations}

   \author{E. Carretti\inst{1}
          }

   \institute{INAF -- Istituto di Radioastronomia, Via Gobetti 101, I-40129 Bologna, Italy
             }

   \abstract{The $B$--Mode of the Cosmic Microwave Background Polarization (CMBP)
promises to detect the gravitational wave background left by Inflation 
and explore this very early period of the Universe.
In spite of its importance, however, the cosmic signal is tiny 
and can be severely limited by astrophysical foregrounds.
In this contribution we discuss about one of the main contaminant, the diffuse
synchrotron emission of the Galaxy. We briefly report about recent
deep observations at high Galactic latitudes,
the most interesting for CMB purposes because of the low emission, and 
discuss the contraints in CMBP investigations. The contamination competes with CMB models with
$T/S = 10^{-2}$--$10^{-3}$, close to the intrinsic limit 
for a 15\% portion of the sky (which is $T/S \sim 10^{-3}$). 
If confirmed by future surveys with larger sky coverage, this gives
interesting perpectives for experiments, that, 
targeting selected low emission regions, could reach this theoretical limit.
}
   \maketitle
%
%________________________________________________________________
%
\section{Introduction}

The polarization of the Cosmic Microwave Background radiation (CMBP) 
is a powerful tool to investigate the early Universe.
The tensor pertubations of the primordial Gravitational Wave (GW) background left by Inflation,
and the reionization history of the Universe can be effectively
studied by the two components the CMBP can be expanded in: the $E$- and $B$-mode
(Zaldarriaga \& Seljak 1997).

\begin{figure}
  \includegraphics[angle=0, width=0.9\hsize]{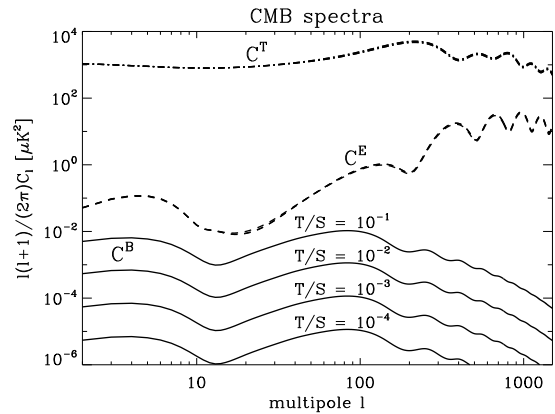}
\caption{Angular power spectra of temperature anisotropy ($C^T$), $E$-mode ($C^E$)
and $B$-mode ($C^B$) of the CMB. Cosmological parameteres of the so-called 
{\it concordance} model after the WMAP data are assumed (\cite{spergel06}~2006). 
In particular an optical depth of the reionized medium of $\tau = 0.1$ is used. 
Spectra for some values of the unknown $T/S$ parameter
are plotted.
\label{cbFig}
}
\end{figure}

The $B$--mode component is faint, but can give us the first tool
to investigate the physics of the Inflation.
In fact, the $B$-mode level on degree scales 
is related to the amount of the primordial GW background emitted by Inflation, which is
usually measured through the tensor-to-scalar
perturbation power ratio $T/S$ (Figure~\ref{cbFig} and, e.g.,
\cite{boyle06}~2006, \cite{kinney06}~2006). 
The $B$-mode spectrum normalization has a linear dependence on it (see Figure~\ref{cbFig}),
while the other CMB components are almost insensitive to these
tensor perturbations\footnote{The temperature spectrum $C^T$ has some
sensitivity for values $T/S > 0.1$.}.
%%%%%%%%%%%%%%%%%%%%%%%%%%%%%%%%%%%%%%%%%%%%%%%%%%%%%%%%%%%%%%%%%%%%%%%%%%%%%
%%%%%%%%%%%%%%%%%%%%%%%%%%%%%%%%%%%%%%%%%%%%%%%%%%%%%%%%%%%%%%%%%%%%%%%%%%%%%
In combination with parameters measured by the CMB Temperature spectrum 
(the scalar pertubation spectral index $n$ and its running $dn / d\ln k$),
it distinguishes among the several Inflation models.

However, $T/S$ is unknown so far 
and only upper limits exist ($T/S < 0.22$, 95\% C.L., \cite{seljak06}~2006).
The many Inflation models predict it to range within orders of magnitude 
(approximately $10^{-4} < T/S < 10^{-1}$),
which corresponds to signals as low as 3--100~nK.
The detection of the $B$-mode is thus fundamental to distinguish among
different models and to measure their relevant parameters, 
like the energy density of the Universe when Inflation itself occurred.
Still undetected, the measurement of this CMBP component
is a hot topic in cosmology.

\section{Contamination by foreground emissions}\label{foregSec}

The weakness of the CMB $B$-mode makes
it easily contaminated by foreground emission from both the Galaxy
and extragalactic sources.
Figure~\ref{cbFig} allows the comparison of the $B$-mode spectrum $C^B$ with the
temperature one ($C^T$). The $B$--mode looks
$3\times 10^3$~--~$1\times 10^4$ times fainter in signal (corresponding to 
$10^5$~--~$10^8$ times in spectrum), depending on the value of $T/S$. 
This way, the higher polarization fraction expected for 
foreground emissions (Galactic emission can be highly polarized, with values of 10--30\% expected for dust and synchrotron emission) makes the polarization case potentially more affected by foreground contamination than the Temperature one.

The study of astrophysical foregrounds is thus crucial
to set the capability of CMBP experiments to investigate the early Universe.
To measure the emission level sets the detectability limit of 
the $B$-mode, and, in turn, the limit of $T/S$. The latter defines which
part of the Inflationary model space is accessible through CMBP.
Also, the weakness of the $B$-Mode signal will most likely require the
application of cleaning procedures to minimize residual contaminations
(e.g. \cite{tegmark00}~2000, \cite{tucci05}~2005, \cite{page06}~2006).
The cleaning technique efficacy is greatly improved by the knowledge of the 
emission properties of the contaminant, like frequency and angular behaviour. 
This further calls for careful foreground characterizations,
especially at high Galactic latitudes, where lowest contamination is expected
and there are better chances to detect the elusive cosmic signal.

At frequencies lower than 100-GHz, the most relevant
contaminant is expected to be the synchrotron emission, that is
best studied at radio wavelengths.
First estimates, based both on total intensity data and polarized emission
on the Galactic plane, depicted a synchrotron emission that severely
contaminates even models with $T/S > 0.1$,
thus compromising $B$--mode investigations even for most optimistic
Inflation scenarios (e.g. \cite{tucci05}~2005).
Direct measurements at high Galactic latitude are thus mandatory 
to do more reliable estimates and set the actual limits on $B$-mode detection
induced by this Galactic foreground component.

\begin{figure}
  \includegraphics[angle=90, width=1.0\hsize]{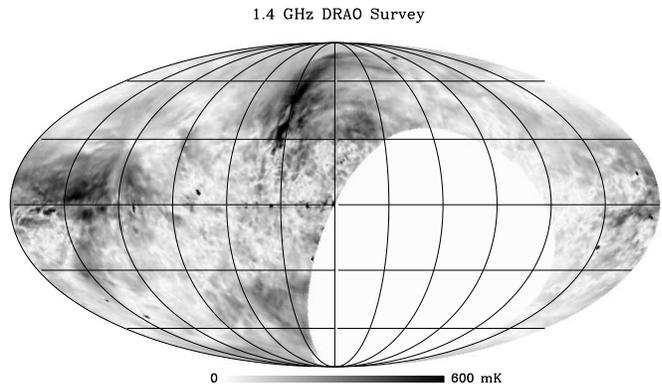}
\caption{Polarized emission map of the DRAO northern sky survey at 
  1.4-GHz (\cite{wolleben06}~2006).
\label{wollebenMapFig}
}
\end{figure}

Because of the frequency behaviour, the Galactic synchrotron is best studied at
radio frequencies, where dominates the other foreground and CMB emissions. 
Polarization measurements with large sky coverage are 
available at 1.4-GHz 
(\cite{wolleben06}~2006, \cite{testori04}~2004).
At this frequency, though, angular power spectra are severely 
modified by Faraday Rotation (FR) effects. 
A power transfer from large to small angular scales can occur, which generates 
depolarization and modifies the shape of the power spectrum  (\cite{carretti05a}~2005a).
This way, 1.4-GHz data are unlikely representative of the Galactic emission
at the higher frequencies of the CMB window (30--150-GHz).
The map of DRAO northern sky
survey 
\footnote{http://www.drao-ofr.hia-iha.nrc-cnrc.gc.ca/26msurvey}
by \cite{wolleben06}~(2006), reported in Figure~\ref{wollebenMapFig}, 
shows evident depolarization in the Galactic
plane area up to a Galactic latitude of $|b| \sim 30^\circ$. After a transition region 
still modified by FR action, only latitudes $|b| > 50^\circ$ looks free from such effects
(see also \cite{carretti05a}~2005a, whose comparison among maps at 5 frequencies in the range
408--1411~MHz highlights such a behaviour).
%%%%%%%%%%%%%%%%%%%%%%%%%%%%%%%%%%%%%%%%%%%%%%%%%%%%%%%%%%%%%%%%%%%%%%%%%%%%%
%%%%%%%%%%%%%%%%%%%%%%%%%%%%%%%%%%%%%%%%%%%%%%%%%%%%%%%%%%%%%%%%%%%%%%%%%%%%%

Deep observations either at higher frequencies or at latitudes
above $|b|> 50^\circ$, and with sensitivity matching CMBP needs,
are thus necessary to reliably estimate the contamination by Galactic
synchrotron emission.

\section{Galactic synchrotron emission at high Galactic latitudes}\label{synchSec}

First deep observations at high Galactic latitudes have been carried out 
in low emission areas. The goal was to first detect
the synchrotron signal at high latitudes, leading 
to survey small fields ($\leq 10$~deg$^2$) to reach proper pixel sensitivity.
Areas in the target fields of BOOMERanG (\cite{masi05}~2005), BaR-SPOrt (\cite{cortiglioni03}~2003), and DASI experiment (\cite{leitch05}~2005) 
have been surveyed with the Parkes, Effelsberg, and ATCA telescope, respectively.
Full details of such observations and description of the emission properties of these areas
are reported in \cite{carretti05b}~(2005b), \cite{carretti06}~(2006), and 
\cite{bernardi06}~(2006), respectively, while the main features are reported in Table~\ref{areasTab}.
%%%%%%%%%%%%%%%%%%%%%%%%%%%%%%%%%%%%%%%%%%%%%%%%%%%%%%%%%%%%%%%%%
\begin{table*}
 \centering
  \caption{Main features of the radio frequency observations in the three low 
           emission areas within the 
           target fields of the BOOMERanG, BaR-SPOrt and DASI experiment, respectively.
           Both the two fields of DASI have been surveyed.}
  \begin{tabular}{@{}lccc@{}}
 \hline
  Area observed                      & BOOMERanG & BaR-SPOrt & DASI \\
 \hline
  Central frequency [GHz]            &  2.332  &  1.402   &  1.380  \\
  FWHM                               & $8.8'$  &  $9.35'$ &  $3.4'$\\
  Galactic Coord.       & $l \sim 255^\circ$, $b \sim -38^\circ$ 
                        & $l \sim 172^\circ$, $b \sim +63^\circ$ 
                        & Field 1:  $l \sim 325^\circ$, $b \sim -58^\circ$ \\
                        & 
                        &  
                        & Field 2:  $l \sim 309^\circ$, $b \sim -62^\circ$ \\
  Area size             & $2.0^{\circ}\times 2.0^{\circ}$
                        & $3.2^{\circ}\times 3.2^{\circ}$ 
                        & $2.0^{\circ}\times 2.0^{\circ}$ (each)\\
  $Q$, $U$ sensitivity in a beam pixel 
                        & 0.27-mK  &  0.86-mK &  3.2-mK\\
  Telescope             & Parkes   & Effelsberg & ATCA\\
  Ref.                  & Carretti et al. 2005b   
                        & Carretti et al. 2006 
                        & Bernardi et al. 2006\\
  \hline
  \end{tabular}
 \label{areasTab}
\end{table*}
%%%%%%%%%%%%%%%%%%%%%%%%%%%%%%%%%%%%%%%%%%%%%%%%%%
These areas are located in low total emission regions,
that cover about 15\% of the sky and are expected to
feature lowest contamination even in polarization.

\begin{figure}
\centering
  \includegraphics[angle=0, width=1.0\hsize]{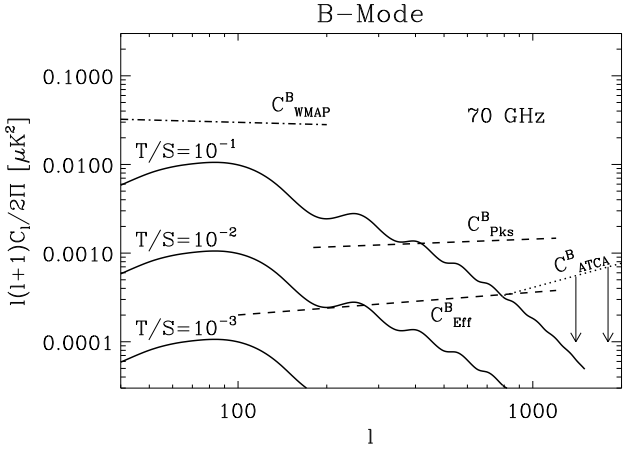}
\caption{$B$--mode power spectra of the Galactic synchrotron emission estimated at 
         70-GHz compared with the CMB ones. 
         The cases of the two low emission areas in the target fields of 
         the BOOMERanG
         ($C^B_{\rm Pks}$) and BaR-SPOrt experiments ($C^B_{\rm Eff}$) are plotted
         along with the upper limit in the DASI fields ($C^B_{\rm ATCA}$).
         CMB spectra are shown for three different $T/S$ values (solid). 
         The general contamination at high Galactic latitude,
         as estimated from WMAP data, is
         also reported: the spectrum of 74.3\% of the sky at 22.8-GHz is considered 
         (\cite{page06}~2006) and extrapolated to 70-GHz.
\label{specBFig}
}
\end{figure}

DASI and BaR-SPOrt areas are at the Galactic latitude of $b\sim -60^\circ$ and 
$b\sim +63^\circ$, respectively, above the limit of $|b|\sim 50^\circ$ 
under which the FR effects significantly modify measurements
at 1.4-GHz. The area in the BOOMERanG field, instead, is at lower latitude 
($b\sim -38^\circ$) and has required observations at
2.3-GHz.
There are indications those observations are actually 
unaffected by FR effects, so that angular power spectra obtained from them
provide first reliable estimates of the synchrotron contamination.
Figure~\ref{specBFig} reports the $B$--mode spectra of the synchrotron 
measured in the BOOMERanG and BaR-SPOrt areas extrapolated
to 70-GHz, that is considered in the best frequency window for CMB 
(\cite{carretti05b}~2005b, \cite{page06}~2006). 
%%%%%%%%%%%%%%%%%%%%%%%%%%%%%%%%%%%%%%%%%%%%%%%%%%%%%%%%%%%%%%%%%%%%%%%%
%%%%%%%%%%%%%%%%%%%%%%%%%%%%%%%%%%%%%%%%%%%%%%%%%%%%%%%%%%%%%%%%%%%%%%%%%%%%%%%%%%%%%%%%%%%%%
The comparison with CMBP spectra for different $T/S$ values 
suggests that the properties of Inflation models with a GW amount
down to $T/S = 10^{-2}$--$10^{-3}$ would be accessible in these low emission
regions,
even without applying any foreground cleaning algorithm. Cleaning procedures
could assure to reach even lower values ($T/S < 10^{-3}$).
The measurements in the DASI fields did not allow any signal detection, even though
an upper limit have been set (\cite{bernardi06}~2006). This has been reported in 
Figure~\ref{specBFig} assuming a frequency slope $\beta = -3.1$ (\cite{bernardi04}~2004).
Although covering multipoles $\ell >800$ far from the angular scales of interest ($\ell \sim 100$), this
upper limit supports the low contamination level resulted 
in the other two fields.

The contribution of the ISM dust has to be considered as well
to evaluate the limits induced by the whole Galactic emission. 
No reliable measurements exist at high Galactic latitude
of its polarized component. \cite{carretti06}~(2006) estimate it 
from the total intensity measurements at high Galactic latitudes (\cite{masi05}~2005)
and assuming the 10~per~cent polarization fraction deduced by
\cite{benoit04}~(2004). They find the dust contribution is similar to the
synchrotron one around 70-GHz and does not significantly change the
$T/S$ limit discussed above.

The recently released 3-years WMAP data depict a different situation for the {\it normal}
high Galactic latitudes conditions. In fact, WMAP has provided an
all-sky polarization map at 23-GHz that makes possible to derive 
the mean emission at high Galactic latitude (\cite{page06}~2006). 
Figure~\ref{specBFig} reports the power spectrum they measured for the
high Galactic latitudes (74.3\% of the sky) 
scaled up to 70-GHz using a frequency slope $\beta = -3.1$.
The CMB signal looks contaminated, with the Galactic
signal competing with even optimistic models with $T/S\sim 0.3$, 
frustrating the possibility to detect the cosmic signal unless high $T/S$ 
values (but disfavoured by last upper limits) or aggressive cleaning procedures.

Therefore, high Galactic latitudes are {\it normally} severely contaminated, 
making even more important to characterize the lowest emission regions.
WMAP missed to detect the signal in such regions 
($S/N < 1$ on angular scale of $4^\circ$), which, conversely, 
we successfully observed (our areas are located in the right side
of the good regions identified by WMAP). Here, as discussed above,
$T/S$ limits are more optimistic and open new and interesting
perspectives, showing the possibility to access a large part
of the Inflation model space.

\section{Discussion}

The results reported in Sect.~\ref{synchSec} depict a situation with significant
contamination even at high Galactic latitudes, but with 
better conditions in the low emission regions. 
Such {\it low lands} represent about 15\% of the sky and
could be the right place where to conduct deep CMBP observations
to look for the $B$-mode.

To limit observations in small regions, however, imposes
intrinsic limitation on the minimum detectable $T/S$, 
mainly because of the leakage from $E$- into the weaker $B$-mode.
In fact, \cite{amarie05}~(2005) find that an all-sky survey would allow a detection 
of $T/S$ with a theoretical sensitivity 
limit\footnote{i.e. in the ideal case of negligible instrument noise.}
of $\Delta T/S = 1.5 \times 10^{-5}$, that becomes
$\Delta T/S = 3.2 \times 10^{-5}$ when 70\%
of the sky is available, $\Delta T/S = 10^{-3}$ for 15\%, and $\Delta T/S = 10^{-2}$ 
in 1\% (3-$\sigma$ C.L.). 
It is worth noting that the class of Inflation models with minimal fine-tuning 
have $T/S$ values ranging within $10^{-3}$ and $10^{-1}$ (\cite{boyle06}~2006), 
for which a 15\% sky portion would be large enough for the first detection of the tensor
CMBP component.

%%%%%%%%%%%%%%%%%%%%%%%%%%%%%%%%%%%%%%%%%%%%%%%%%%%%%%%%%%%%%%%%%%%%%%%%
\begin{figure}
\centering
  \includegraphics[angle=0, width=0.7\hsize]{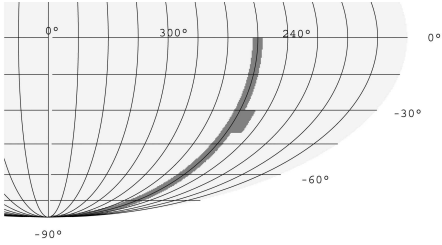}
\caption{The PGMS survey coverage in Galactic coordinates.
         \label{stripFig}
}
\end{figure}
%%%%%%%%%%%%%%%%%%%%%%%%%%%%%%%%%%%%%%%%%%%%%%%%%%%%%%%%%%%%%%%%%%%%%%%%
%%%%%%%%%%%%%%%%%%%%%%%%%%%%%%%%%%%%%%%%%%%%%%%%%%
\begin{table}
 \centering
  \caption{Main features of the PGMS survey.}
  \begin{tabular}{@{}lr@{}}
  \hline
  Central frequency                  &  2.332-GHz \\
  Bandwidth                & $128$-MHz \\
  FWHM                               & $8.8'$ \\
  Galactic longitude                 & $l \sim 255^\circ$ \\
  Area size                          & $5.0^{\circ}\times 90.0^{\circ}$ \\
  $Q$, $U$ sensitivity in a beam pixel & 0.5-mK \\
  Telescope                          & Parkes \\
  \hline
  \end{tabular}
 \label{pgmsTab}
\end{table}
%%%%%%%%%%%%%%%%%%%%%%%%%%%%%%%%%%%%%%%%%%%%%%%%%%

Despite of these limitation, the minimum $T/S$ value detectable in 15\% of the
sky almost matches the limit imposed by the foregrounds in low emission regions.
The search for the $B$-mode in a sky portion of such a size could thus be a good
trade-off between instrinsic and foregrounds limits.
In addition, the weakness of the $B$-mode signal makes already a challenge
to detect the signal for $T/S=0.1$ with the present technology (\cite{cortiglioni06}~2006). 
It is likely that significant technological improvements will be necessary
before scientists can face an all-sky mapping mission with a sensitivity 
able to match the intrinsic limit of $\Delta(T/S)\sim 10^{-5}$.
Therefore, an experiment aiming at detecting the $B$-mode in a smaller region 
(10-15\% of the sky) can be an interesting intermediate step that would allow us to
probe Inflation models down to $T/S = 10^{-3}$.

The observations conducted in the {\it low lands} cover small areas. 
Although taken in three independent samples,
surveys of large portions of the sky are necessary 
to understand whether the areas observed so far are peculiar {\it lucky} 
lowest emission cases or are representative of the
conditions of the {\it low lands}.
New information are expected to come from the recently completed 
all-sky mapping at 1.4-GHz, even if, as mentioned in Sect.~\ref{foregSec},
these are likely modified by the FR action up to $|b|\sim50^\circ$.
First analyses using these data have been reported by \cite{laporta06}~(2006). 
However, they regard half a sky and contain both large local high emission regions 
and the Galactic disc that is strongly modified by FR.

Deeper and higher frequency data will come soon from the Parkes Galactic
Meridian Survey (PGMS, \cite{carretti05c}~2005c)
aimed at surveying a Galactic meridian at 2.3-GHz (Figure~\ref{stripFig}).  
The survey is in progress at the Parkes radiotelescope and main features
are reported in Table~\ref{pgmsTab}.
It will explore the polarized Galactic diffuse emission along the
meridian $l=255^\circ$ from Galactic plane to south Pole.
The frequency is high enough to avoid significant FR effects and main goals
are to reveal
the intrinsic Galactic emission and study its behaviour with the Galactic
latitude. This meridian goes through one of the low emission regions 
visible in the WMAP data and more extended information about the
Galactic emission level in the {\it low lands} are expected.

\begin{acknowledgements}
The author thanks Stefano Cortiglioni for discussions and suggestions.
Some of the results in this paper have been derived using the 
HEALPix package (http://healpix.jpl.nasa.gov,
\cite{gorski05}~2005).
We aknowledge the use of the CMBFAST package. The Dominion Radio Astrophysical Observatory is operated as a national Facility by the National Research Council Canada.
\end{acknowledgements}

\end{document}